\documentclass[epj]{svjour}
\usepackage{amsfonts,amssymb,amsmath,hyperref,graphicx,epsfig}
\newcommand{\ket}[1]{| #1 \rangle}
\newcommand{\bra}[1]{\langle #1 |}
\newcommand{\avg}[1]{\langle #1 \rangle}
\begin{document}
\title{Interacting heavy fermions in a disordered optical lattice}
\author{Bo-Nan Jiang\inst{1,4}
\and Jun Qian\inst{1} \fnmsep \thanks{\email{jqian@mail.siom.ac.cn}}
\and Wen-Li Wang\inst{1,3}
\and Juan Du\inst{2} \fnmsep \thanks{\email{dujuan@mail.siom.ac.cn}}
\and Yu-Zhu Wang\inst{1} \fnmsep \thanks{\email{yzwang@mail.shcnc.ac.cn}}
}
\institute{Key Laboratory for Quantum Optics, Shanghai Institute of Optics and Fine Mechanics, Chinese Academy of Sciences, Shanghai 201800, China
\and
State Key Laboratory of High Field Laser Physics, Shanghai Institute of Optics and Fine Mechanics, Chinese Academy of Sciences, Shanghai 201800, China
\and
State Key Laboratory of Precision Measurement, East China Normal University, Shanghai 200241, China
\and
University of Chinese Academy of Sciences, Beijing 100049, China
}
\date{Received: date / Revised version: date}
\abstract{
We have theoretically studied the effect of disorder on ultracold alkaline-earth atoms governed by the Kondo lattice model in an optical lattice via simplified double-well model and hybridization mean-field theory. Disorder-induced narrowing and even complete closure of hybridization gap have been predicted and the compressibility of the system has also been investigated for metallic and Kondo insulator phases in the presence of the disordered potential. To make connection to the experimental situation, we have numerically solved the disordered Kondo lattice model with an external harmonic trap and shown both the melting of Kondo insulator plateau and an compressibility anomaly at low-density.
}
\maketitle

\section{Introduction}
\label{sec:Intro}
Though Anderson localization describes the disordered noninteracting particles well \cite{Anderson1958}, understanding the interplay between interaction and disorder in quantum many-body systems has still been a challenging topic for condensed matter physics community \cite{Trivedi1997}. Cheerfully, recent developments of ultracold gases in laser speckles or incommensurate optical lattices provide a new platform with both experimentally controlled interaction and disorder to reveal the physics of noninteracting and interacting particles in disordered potentials \cite{Billy2008,Roati2008,Inguscio2010,DeMarco2009,DeMarco2010,Lewenstein2010}. Nowadays, though the task of mimicking quantum magnetism has still been hindered by the required low temperature of spin ordering \cite{Esslinger2013a}, the milestone realizations of the fermionic Mott insulator (MI) for alkali or alkaline-earth atoms (AEAs) \cite{Esslinger2008,Schneider2008,Kyoto2011,Kyoto2012} have inspired the blossom of experimental and theoretical studies on disordered strongly correlated Fermi gases. Experimentally, the transition from superfluid to a glasslike phase has been identified in disordered strongly interacting $^6$Li$_2$ molecules \cite{Brantut2012,Esslinger2013b}, and the disruption of the Hubbard gap by disorder-induced density fluctuations has been probed in MI regime with ultracold fermionic $^{40}$K$ $ atoms \cite{DeMarco2013}. Theoretically, the real-time dynamics of strongly correlated three-dimensional Fermi system in disordered potentials has been investigated combining time-dependent density-functional theory and dynamical mean-field theory \cite{Kartsev2013}, and the ground-state properties of the disordered one-dimensional Fermi-Hubbard model with harmonic confining potential have also been studied within a density-functional scheme \cite{Gao2006}, where the validity of local density approximation (LDA) is limited to weak \cite{Gao2014,Gao2011} or long-range correlated \cite{Thomas2012} disorder.

As another most active field in condensed matter physics, heavy fermion compounds can be modeled as a lattice of localized spin moments coupled to a band of conduction electrons via Kondo-exchange interaction \cite{Hewson1993,Ueda1997,Si2010}. Recently, Gorshkov \textit{et al.} and Foss-Feig \textit{et al.} have shown that fermionic AEAs have unique properties that allows for reconstructing the Kondo lattice model (KLM) and investigating heavy fermion physics from different perspectives \cite{Gorshkov2010,Rey2010a,Rey2010b}, which has been strongly supported by the very recent experimental observation of remarkable two-orbital spin-exchange interactions in both $^{87}$Sr$ $ \cite{Ye2014} and $^{173}$Yb$ $ \cite{Bloch2014}. To the best of our knowledge, the zero-temperature properties of interacting heavy fermions in a disordered optical lattice have not been uncovered yet \cite{Lewenstein2010}. To implement the disordered KLM with the AEAs, ultracold atoms in two clock states $\ket{^1$S$_0}$ ($\ket{g}$) and $\ket{^3$P$_0}$ ($\ket{e}$) are trapped in two independent optical lattice potentials at the same periodicity \cite{Daley2008}, where mobile $g$ atoms (conduction electrons) coexist with a MI background of $e$ atoms (localized spin moments). The disordered potential is created by a state-dependent external laser speckle radiating the itinerant $g$ atoms and the strength of the disorder is characterized by the average speckle potential energy at the focus of the lens, which can be adjusted by tuning the speckle laser power.

In this article, the physics of interacting heavy fermions in a disordered optical lattice has been captured both qualitatively in the double-well (DW) model by analytic solutions and quantitatively by hybridization mean-field theory (hMFT) in the thermodynamic limit. From the theoretical side, we extend the arithmetically averaging approach in Ref. \cite{Krutitsky2006} to the heavy-fermion system. Subsequently, a LDA+hMFT calculation is performed to incorporate harmonic confinement effect following Ref. \cite{Bissbort2009,Bissbort2010}.
\section{The double-well model}
\label{sec:DW}
\begin{figure}
\includegraphics[width=1\linewidth]{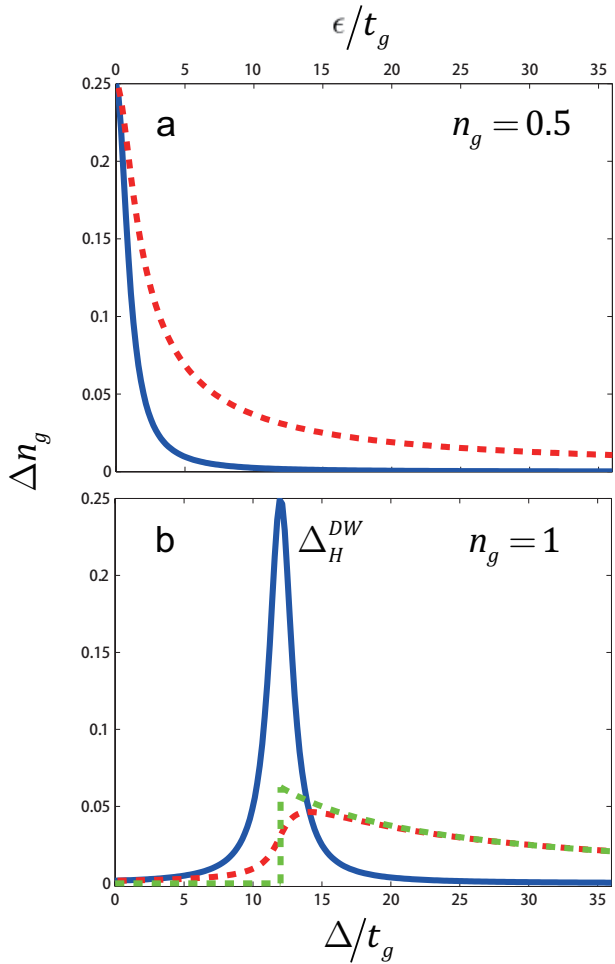}
\caption{Density fluctuation of the ground states of the half-filling ($a$) and unit-filling ($b$) KLM in tilted (blue solid) and randomly tiled (red dashed) DW potentials. The green dashed line shows the density fluctuation in the unit-filling case calculated based on the simplified $\delta$-function lineshape of the resonant tunneling.}
\label{fig:DW}
\end{figure}
Besides the sophisticated numerical recipes, such as density matrix renormalization group, the DW models can also reveal the basic physics of a disordered quantum gas qualitatively, especially the interplay between interaction and disorder in synthesis disordered systems \cite{Zhou2010}. The model we propose describes interacting heavy fermions in a disordered DW potential
\begin{eqnarray}
\label{eq:DW}
H_{DW} &=& -t_g \sum_{\sigma\in\uparrow,\downarrow}(c^{\dagger}_{Lg\sigma}c^{}_{Rg\sigma}+h.c.) + J_K \sum_{i\in L, R}\vec{S}_{ig}\cdot\vec{S}_{ie} \nonumber\\
       &+& \frac{\epsilon}{2}(n_{Lg}-n_{Rg}),
\end{eqnarray}
where $c^{\dagger}_{i\alpha\sigma}$ ($c_{i\alpha\sigma}$) creates (destroys) one $\alpha\in\{e,g\}$ atom at spin state $\sigma$ on site $i$. $\vec{S}_{i\alpha}=\frac{1}{2}\sum\limits_{\sigma,\sigma'}c^{\dagger}_{i\alpha\sigma}
\vec{\sigma}_{\sigma\sigma'}c_{i\alpha\sigma'}$ ($\vec{\sigma}$ is the vector of Pauli matrices). $t_g$ and $J_K$ are magnitudes of tunneling and Kondo-exchange interaction respectively. And $\epsilon$ measures the energy level mismatch between the tilted wells. In the DW model, disorder is mimicked by a random tilting following box distribution $p(\epsilon)=\Theta(\Delta-|\epsilon|)/2\Delta$ between two wells, with the disorder strength parameter $\Delta$. For a single-site operator $O$, ${\avg{O}}_\epsilon$ is the ground state observation at a fixed tilting $\epsilon$, and the average expectation for a variety of random samples is given by
\begin{equation}
\overline{\avg{O}}_{\Delta}=\int_{-\Delta}^\Delta d\epsilon p(\epsilon){\avg{O}}_\epsilon.
\end{equation}
Here, to approach the effect of disorder on the conducting and insulating ground states in the DW model, we focus on the half-filling ($n_g=0.5$) and unit-filling ($n_g=1$) cases respectively.

Considering the experimental accessibility, the requirement to reach the temperature smaller than the Kondo temperature $T_K\sim t_ge^{-t_g/J_K}$ can always be relaxed by choosing a larger $J_K$ in optical lattices \cite{Rey2010b}. In this article, we choose $J_K=8t_g$ and obtain $T_K\sim t_g$, so only a temperature smaller than the tunneling energy is needed to realize the artificial heavy-fermion system. Moreover, in the chosen strong-coupling parameter regime, we can both restrict the composite object of a pair of $g$ and $e$ atom on the same DW site to Kondo singlet and apply the hybridization mean-field treatment in the thermodynamic limit reasonably.

Approximately, the DW with half-filling of $g$ atoms in the strong-coupling limit describes a single Kondo singlet tunnels between subwells with effective tunneling amplitude $t_g/2$ in the representation consisting of holon state $\ket{0}$ and singlet state $\ket{s}$ \cite{Ueda1991a,Ueda1991b}. In Fig. \ref{fig:DW}$a$, the ground state continuously transforms from the extended state $1/\sqrt{2}(\ket{s,0}+\ket{0,s})$ to the localized state $\ket{0,s}$, since the energy mismatch $\epsilon$ reduces the tunneling of Kondo singlet between the subwells. And the density fluctuation of the tilted DW model
\begin{equation}
\avg{\Delta n_g}_{\epsilon}=\left[\frac{t_g(|\epsilon|+\sqrt{t^2_g+\epsilon^2})}
{t^2_g+(|\epsilon|+\sqrt{t^2_g+\epsilon^2})^2}\right]^2
\end{equation}
decreases with the increasing energy mismatch. Thus, when disorder introduces random relative energy mismatches, the density fluctuation
\begin{equation}
\overline{\avg{\Delta n_g}}_{\Delta}=\frac{t_g\arctan{\frac{\Delta}{t_g}}}{4\Delta}
\end{equation}
also decreases monotonically, as the weight of the localized state increases with the disorder strength. Accordingly, in a strongly disordered dilute heavy-fermion gas which can be interpreted as being composed of holon sites and half-filling DWs approximately, $g$ atoms tend to be localized in the randomly distributed low-energy subwells in the disorder landscape. And this explains the compressibility anomaly at low density qualitatively \cite{Xie2002}.

The Hilbert space of the unit-filling DW model in the strong-coupling limit \cite{Jurecka2001} consists of the Kondo insulating state $\ket{s,s}$ protected by the DW gap $\Delta^{DW}_{H}=3/2J_K$ and the dipole state $\ket{0,D}$ describing a holon-doublon pair on nearest-neighbor sites. In Fig. \ref{fig:DW}$b$, when such a DW system is tilted, the Kondo insulating state and the dipole state become degenerate and coupled resonantly by the $g$ atom tunneling at $\epsilon=\Delta^{DW}_{H}$, where the ground state $1/\sqrt{2}(\ket{s,s}+\ket{0,D})$ is most conductive. And the density fluctuation
\begin{equation}
\avg{\Delta n_g}_{\epsilon}=\left[\frac{t_g\left[(|\epsilon|-\Delta^{DW}_{H})+\sqrt{t^2_g+(|\epsilon|-\Delta^{DW}_{H})^2}\ \right]}
{t^2_g+\left[(|\epsilon|-\Delta^{DW}_{H})+\sqrt{t^2_g+(|\epsilon|-\Delta^{DW}_{H})^2}\ \right]^2}\right]^2
\end{equation}
manifests the corresponding resonant lineshape. The resonant feature of density fluctuation vanishes when a disordered DW potential tunes the energy difference between the Kondo insulating state and the dipole state randomly. Being understood within a simplified picture where a $\delta$ function describes the resonant tunneling probability \cite{Cohen1998}, the Kondo insulating state preserves till the disorder strength reaches a discontinues singularity at $\Delta=\Delta^{DW}_{H}$, and then the probability of the resonant tunneling drops following the power law $\propto1/\Delta$ as strong disorder prefers the dipole deformed ground state. Obviously, the effect of disorder on the resonant tunneling (also density fluctuation and compressibility) changes from enhancement to suppression at $\Delta=\Delta^{DW}_{H}$ where the disorder averaged coupling between the Kondo insulating state and the dipole state is strongest. In the actual DW system, though the $g$ atom tunneling broadens the simplified $\delta$-function lineshape of the resonant tunneling, the density fluctuation
\begin{equation}
\overline{\avg{\Delta n_g}}_{\Delta}=\frac{t_g\left[\mathrm{arccot}\frac{t_g}{\Delta-\Delta^{DW}_{H}}+\arctan{\frac{\Delta^{DW}_{H}}{t_g}}\right]}{4\Delta}
\end{equation}
manifests the similar trend as the simplified picture except the analytic ramping-up behavior around $\Delta=\Delta^{DW}_{H}$.

In the thermodynamic limit, the resonant family is supplemented with multiple dipole deformations of the Kondo insulator \cite{Sachdev2002}, and a resonant excitation probability versus $\epsilon$ between neighboring lattice sites in a clean system is expected as its Mott counterpart \cite{Greiner2002}. Since only separate creations of resonant dipole states on nearest-neighbor links are possible \cite{Sachdev2002}, the DW model may still provide a reasonable qualitative understanding for the effect of disorder on Kondo insulators with the chemical potential lying in the middle of the hybridization gap.
\section{Hybridization mean-field theory}
\label{sec:HMFT}
\begin{figure}
\includegraphics[width=1\linewidth]{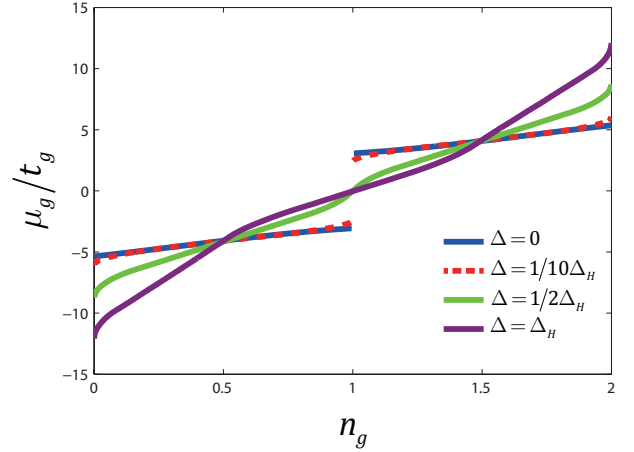}
\caption{The effect of disorder on hybridization gap structure. And the closure of the hybridization gap has been identified at $\Delta=\Delta_H/2$.}
\label{fig:gap}
\end{figure}
\begin{figure}
\includegraphics[width=1\linewidth]{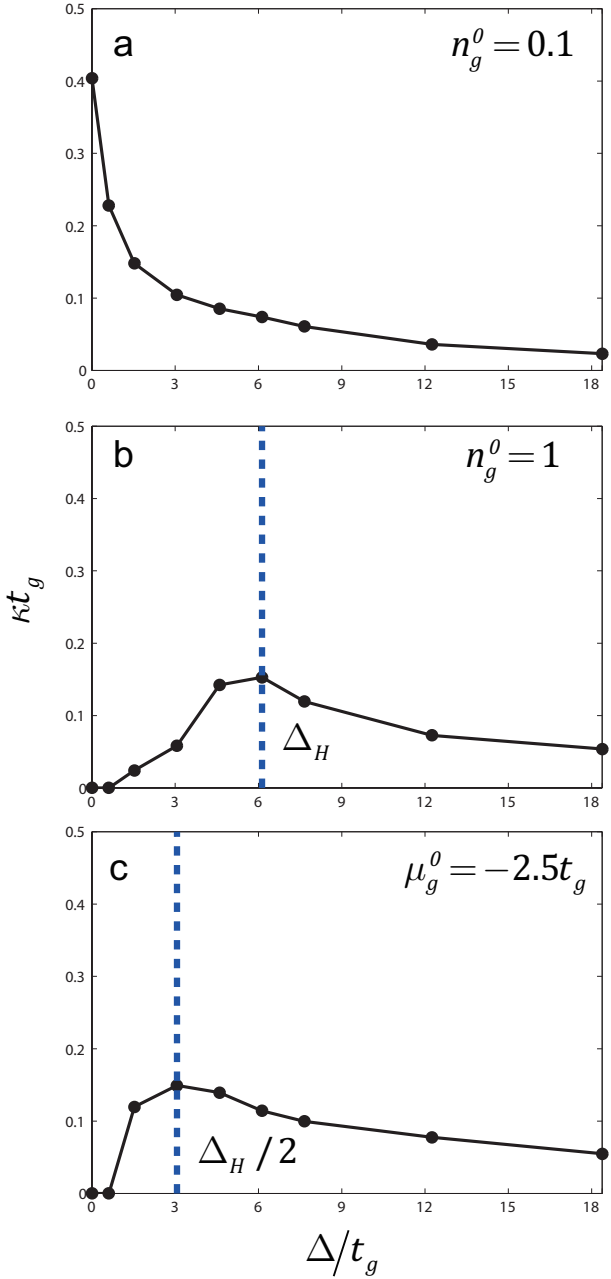}
\caption{The effect of disorder on compressibility in dilute metallic phase ($a$) and Kondo insulator phases with the chemical potential lying in the middle ($b$) and near the edge ($c$) of the hybridization gap.}
\label{fig:compressibility}
\end{figure}
In the thermodynamic limit, the disordered heavy-fermion lattice gas is governed by the single-band disordered KLM Hamiltonian
\begin{eqnarray}
\label{eqnDKLM}
	H_{dKLM} &=& -t_g\sum_{\left<i,j\right>,\sigma} c^{\dagger}_{ig\sigma}c^{}_{jg\sigma} + J_K \sum_{i}\vec{S}_{ig}\cdot\vec{S}_{ie} \nonumber\\
		 	 &+& \sum_{i}(\epsilon_i-\mu_{g})n_{ig} +\sum_{i}(n_{ie}-1)\mu_{e},
\end{eqnarray}
where the onsite energies $\epsilon_i$ of diagonal uncorrelated disorder are assumed to be distributed according to $p_\epsilon=\Theta(\Delta-|\epsilon|)/2\Delta$ with the strength of disorder $\Delta$. $\mu_{g}$ and $\mu_{e}$ are chemical potentials of $g$ and $e$ atoms respectively. And the last term is the constraint of the unit filling of the MI background of $e$ atoms.

As having been verified in the heavy-fermion materials such as YbFe$_{4}$Sb$_{12}$ and CeRu$_4$Sb$_{12}$ \cite{Dordevic2001}, the hybridization between $g$ and $e$ atoms can be treated as the order parameter
\begin{equation}
\eta_i=\frac{1}{2}\sum_{\sigma}\langle c^{\dagger}_{ig\sigma}c_{ie\sigma}+c^{\dagger}_{ie\sigma}c_{ig\sigma}\rangle,
\end{equation}
in the picture of which, disorder fluctuates the hybridization in the heavy-fermion system and induces a bath of the order parameters $\eta_i$ that breaks the translational symmetry of the multiple site lattice model \cite{Bissbort2009}. Consequently, the translationally invariant hMFT can not be simply applied as Ref. \cite{Rey2010b}.

One possible solution of the disorder-induced broken translational symmetry mentioned above is the arithmetically averaging approach in Ref. \cite{Krutitsky2006}, where the disorder-induced fluctuations of the order parameters are accounted for by the disorder averaging $\int^{\Delta}_{-\Delta} d\epsilon p(\epsilon)\dots$, and the corresponding mean-field Hamiltonian can be considered to be translationally invariant approximately. Following Ref. \cite{Krutitsky2006}, the disordered KLM is approximated by an effective translationally invariant mean-field Hamiltonian
\begin{eqnarray}
\label{eqnMF}
H_{MF}&=&-t_g\sum_{\left<i,j\right>,\sigma}c^{\dagger}_{ig\sigma}c^{}_{jg\sigma}-\frac{J_K}{2}\overline{\eta}
\sum_{i,\sigma}(c^{\dagger}_{ig\sigma}c_{ie\sigma}+c^{
\dagger}_{ie\sigma}c_{ig\sigma})\nonumber\\
&+&\frac{J_K}{2}\mathcal{N}\overline{\eta}^2+(\epsilon-\mu_g)\mathcal{N}n_g+\mathcal{N}(n_{e}-1)\overline{\mu}_{e},
\end{eqnarray}
where $\overline{\eta}=\frac{1}{2}\sum\limits_{\sigma} \int^{\Delta}_{-\Delta} d\epsilon p(\epsilon) \langle c^{\dagger}_{ig\sigma}c_{ie\sigma}+c^{\dagger}_{ie\sigma}c_{ig\sigma}\rangle(\epsilon)$ and $\overline{\mu}_{e}=\int^{\Delta}_{-\Delta} d\epsilon p(\epsilon) \mu_e(\epsilon)$. $\mathcal{N}$ is the total number of lattice sites. $\overline{\eta}^{gs}$ and $\overline{\mu}_e^{gs}$ of the disordered heavy-fermion system's ground state are determined by the the self-consistency condition and the unit-filling constraint \cite{Rey2010b}
\begin{eqnarray}
	\frac{\partial{\overline{E}(\overline{\eta},\overline{\mu}_e)}}{\partial{\overline{\eta}}}=&0&, \nonumber \\
	\frac{\partial{\overline{E}(\overline{\eta},\overline{\mu}_e)}}{\partial{\overline{\mu}_e}}=&0&,
\end{eqnarray}
where the disorder averaged total energy per site $\overline{E}(\overline{\eta},\overline{\mu}_e)=\int^{\Delta}_{-\Delta} d\epsilon p(\epsilon) \frac{\bra{\overline{\eta},\overline{\mu}_e,\epsilon}H_{MF}\ket{\overline{\eta},\overline{\mu}_e,\epsilon}}{\mathcal{N}}$. Subsequently, any disorder averaged ground state observation of the single-site operator $O$ can be expressed reasonably as \cite{Krutitsky2006}
\begin{equation}
\label{eqavg}
\overline{\avg{O}}_{\Delta}=\int^{\Delta}_{-\Delta} d\epsilon p(\epsilon) \bra{\overline{\eta}^{gs},\overline{\mu}_e^{gs},\epsilon}O\ket{\overline{\eta}^{gs},\overline{\mu}_e^{gs},\epsilon}.
\end{equation}
Note that in Ref. \cite{Krutitsky2006}, the arithmetically averaging approach fails to describe the Bose glass phase in the disordered Bose-Hubbard model at zero temperature and any finite tunneling amplitude \cite{Bissbort2009}. The invalidity of this approach is caused by the decoupling mean-field approximation applied to the nearest-neighbor tunneling term, which describes the strength of the coupling to the nearest-neighbor sites by the disorder averaged order parameter. However, in our work, the disordered heavy-fermion system is treated by the hybridization mean-field theory that uses onsite hybridization order parameter $\eta_i$ and applies no mean-field approximation to the nearest-neighbor tunneling term. As a result, though we apply the arithmetically averaging to describe the disorder-induced fluctuations of $\eta_i$ and reach the translationally invariant hybridization mean-field Hamiltonian with the disorder averaged hybridization order parameter $\overline{\eta}$, the invalidity caused by the decoupling mean-field approximation mentioned above shall evidently not be inherited by the hybridization mean-field Hamiltonian with the exact form of the nearest-neighbor tunneling.

In Fig. \ref{fig:gap}, the hybridization gap shrinks with the increasing disorder strength and completely destroyed at strong disorder. In the clean system, the chemical potential $\mu_g$ of the heavy fermion system is a particle-hole symmetric function of the $g$ atom density $n_g$, whose discontinuous singularity at unit filling stands for the hybridization gap. And Kondo insulator phase forms when the chemical potential of ground state lies in the hybridization gap. Note that hMFT underestimates the clean hybridization gap by predicting $\Delta_H\sim3/4J_K$ in the chosen parameter regime, for exact results in the strong-coupling limit give $\Delta_{H}=3/2J_K$ \cite{Rey2010a}. Once the disorder potential is introduced, the holon and doublon sites emerge, neither of which have onsite Kondo-exchange interaction or interband hybridization. Thus, the hybridization gap of the disordered KLM narrows as the disorder-induced creation of holon and doublon sites reduces interband hybridization. When the disorder strength reaches $\Delta_H/2$, the hybridization gap disappears completely and the system can possess no insulating phases at all.

On the other hand, the compressible metallic phase and incompressible insulating phases of the disordered KLM can be characterized by compressibility $\kappa=\overline{\avg{1/(\frac{\partial\mu_g}{\partial n_g})|_{n^0_g}}}_{\Delta}$ or $\overline{\avg{(\frac{\partial n_g}{\partial\mu_g})|_{\mu^0_g}}}_{\Delta}$, where density $n_g$ or chemical potential $\mu_g$ are independent variables respectively. For the dilute metallic phase with constant filling $n^0_g=0.1$ (Fig. \ref{fig:compressibility}$a$), the metallic phase evolves into a poorly conducting state at strong disorder as usual quantum Fermi gas \cite{Esslinger2013b}. The reason for this phenomenon is that the strong disorder suppresses the nearest-neighbor tunneling significantly and dilute $g$ atoms tend to occupy the low-energy wells in the disorder landscape \cite{Gao2006}. The dropping of compressibility shows clearly that the dilute heavy-fermion system stiffens remarkably with increasing disorder strength. Differently, in the disordered potential, the Kondo insulator phase with the chemical potential lying in the middle of the hybridization gap (Fig. \ref{fig:compressibility}$b$) has zero chemical potential and is constantly unit-filling as the hybridization gap narrows under disorder (see Fig. \ref{fig:gap}). This insulating phase is robust at weak disorder for dipole states are hardly excited. With the increasing disorder strength, the compressibility of the heavy-fermion system is enhanced gradually by the disorder-induced coupling to the single or multiple dipole deformation. While in strong disorder regime, the compressibility is suppressed since the coupling between the Kondo insulator and the deformation states is reduced. The crossover happens at the critical value $\Delta=\Delta_H$, where the disorder averaged coupling between the Kondo insulating state and the single or multiple dipole states is strongest. It is interesting that we have found a similarity between density fluctuation in the DW model and compressibility in the thermodynamic limit, as shown in Fig. \ref{fig:DW} and Fig. \ref{fig:compressibility}$a-b$. And since the density fluctuation is directly related to the compressibility \cite{Chin2009}, it may indicate the validity of the qualitative DW picture in the disordered heavy-fermion system.

However, in the Kondo insulator phase with the chemical potential locating near the edge of the hybridization gap, we choose constant chemical potential $\mu^0_g=-2.5t_g$ and leave the $g$ atom density of the ground state varying with disorder strength. Since quantum phases in the heavy-fermion system are filling induced, the changing of $g$ atom density leads to quantum phase transition (QPT) obviously. And understanding of the underlying physics has been obtained from the hybridization gap structure. In Fig. \ref{fig:compressibility}$c$, Kondo insulator manifests nontrivial robustness at weak disorder. The QPT towards the compressible metallic phase with fractional density does not occur till the unit-filling Kondo insulator phase reaches the shrunk hybridization gap edge, which is consistent with the enhancement of compressibility at $\Delta\geq\Delta_H/2+\mu^0_g$. At strong disorder $\Delta>\Delta_H/2$, the hybridization gap is destroyed and the gapless system can possess no incompressible insulating phase (see Fig. \ref{fig:gap}). In such a compressible phase of the gapless system, the compressibility has been suppressed by disorder. And we find the crossover of the effect of disorder from increasing to decreasing compressibility happens at critical value $\Delta=\Delta_H/2$, where disorder is just strong enough to make the hybridization gap vanish. Actually, we find that the above mentioned behavior of compressibility is a common response of the Kondo insulator to the disordered potential. The crossover point of the effect of disorder on compressibility is indeed dependent on the chemical potential $\mu^0_g$ of the Kondo insulator phase, with the critical value of the disorder strength ranging from $\Delta_H/2$ to $\Delta_H$ correspondingly (not shown here).
\section{Incorporating experimental harmonic confinement with LDA}
\begin{figure}
\includegraphics[width=1\linewidth]{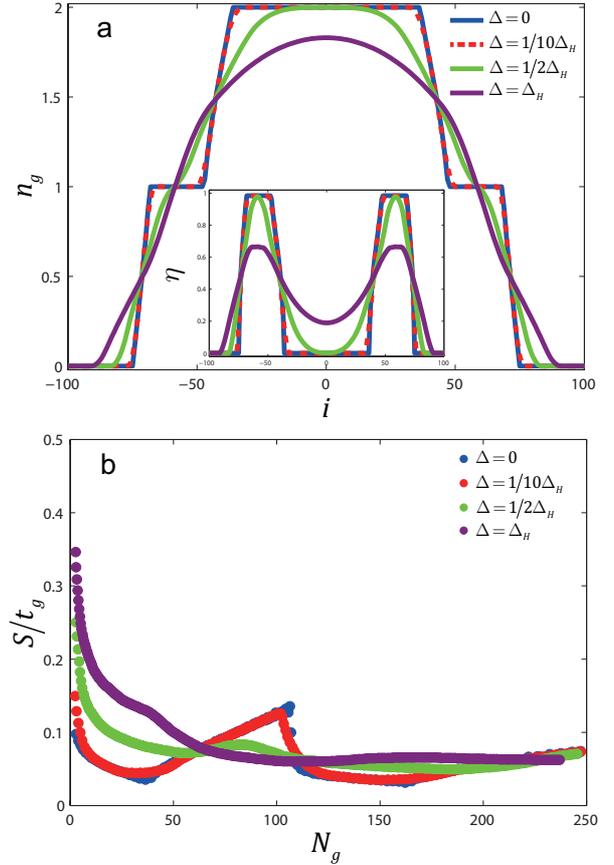}
\caption{The effect of disorder on density profile of the wedding cake structure ($a$) and thermodynamic stiffness ($b$). Inset of ($a$): the corresponding hybridization profile.}
\label{fig:LDA}
\end{figure}
From the experimental side, the strongly correlated lattice gases are usually trapped in the harmonic confining potential and manifest the wedding cake structure, where quantum phases and QPTs are probed with density profile and compressibility \cite{Schneider2008,Chin2009}. However, the arithmetically averaged hMFT in the foregoing section is in nature a single-site theory. To extend the properties of interacting heavy fermions in a disordered optical lattice obtained via hMFT to the experimental situation, the spatial inhomogeneity of the slowly varying $\Omega=2.5t_g/1000$ harmonic confinement $\sum\limits_{i}\Omega i^2n_{ig}$ is treated via LDA \cite{Jaksch1998}, where the trapping potential is considered as a local chemical potential $\mu_{jg}=\mu_g-\Omega j^2$ in the effective hMFT Hamiltonian. And a similar approach combing single-site mean-field theory and LDA has already solved the disordered Bose-Hubbard model in the harmonic trap successfully \cite{Bissbort2010}.

For the density profile in Fig. \ref{fig:LDA}$a$, the wedding cake structure shows two insulator plateaus surrounded by the liquid edges: (1) the Kondo insulator plateau with unit-filling of $g$ atoms has strongest interband hybridization. (2) while the normal insulator plateau consisting of doublon sites experiences neither Kondo-exchange interaction or interband hybridization. The the robustness of the wedding cake structure is obvious at weak disorder, though the slight melting at the edge of the insulating plateau indicates the smoothing of the singularity of the hybridization gap. With increasing disorder strength, the closure of the hybridization gap has been identified when the Kondo insulator plateau vanishes at $\Delta=\Delta_H/2$. The corresponding hybridization profile shows that the disorder-induced density fluctuation \cite{DeMarco2013} disrupts the Kondo insulator plateau consisting of Kondo singlets by reducing the interband hybridization between $g$ and $e$ atoms via creating zero-hybridization holon and doublon sites. At strong disorder, the wedding cake structure has melted completely and manifests the bell-shape density profile consistent with the gapless system. And as the disorder reduces the chemical potential $\mu_g$ of the dilute compressible phase (see Fig. \ref{fig:gap}), the considerable leakage though the edge of the harmonic confinement has also been predicted. Note that the local compressibility of the disordered KLM can also be extracted from the density profile following the definition $\kappa=\overline{\avg{\frac{\partial{n_g(r)}}{\partial{\Omega r^2}}}}_{\Delta}$ in Ref. \cite{Chin2009}. By measuring the local compressibility under different disorder strength, the results in Fig. \ref{fig:compressibility} can be verified experimentally.

Since disorder and interactions affect thermodynamic quantities, like the solid-state systems, the global compressibility is important in understanding the effect of disorder on the wedding cake structure \cite{Ilani2000,Dultz2000}. In Fig. \ref{fig:LDA}$b$, we study the global compressibility by calculating stiffness $S=\overline{\avg{\frac{\partial{\mu_g}}{\partial{N_g}}}}_{\Delta}$, where $N_g$ is the total number of the $g$ atoms \cite{Gao2006,Dultz2000}. The singularities exhibited by the global compressibility indicate the QPTs between metallic and insulating phases accessed by varying $N_g$. As the hybridization gap narrows, the increasing disorder strength smooths the singularities out. And at strong disorder, the compressibility anomaly at low density is very obvious, being consistent with Fig. \ref{fig:compressibility}$a$ and the qualitative DW picture.
\section{Conclusion}
Motivated by the very recent theoretical prediction and experimental observation of two-orbital SU(N) magnetism \cite{Gorshkov2010,Ye2014,Bloch2014}, we have studied the response of ultracold AEAs governed by the KLM to the diagonal uncorrelated disordered potential. By both the DW model and hMFT, besides the transition of heavy fermions from metallic phase to poorly conducting phase at strong disorder as usual quantum Fermi gas, we have demonstrated that for Kondo insulators with the chemical potential lying in the middle and near the edge of the hybridization gap, the crossover of the effect of disorder on compressibility is related to the disorder averaged coupling between the insulating and dipole states and the vanishing of hybridization gap respectively. We have also shown that our demonstration is accessible in the experimental harmonic confining potential by probing the density profile and compressibility with the state-of-the-art experimental techniques. We hope this work can serve as a reasonable guide for the underlying physics of the disordered KLM and contribute to the experimental investigation of the disordered KLM with ultracold AEAs in the future.
\section*{Acknowledgements}
We thank Gao Xianlong and Michael Foss-Feig for fruitful discussions. The work is supported by grants from the National Natural Science Foundation of China (No. 11104292) and the National Basic Research Program of China (No. 2011CB921504). J.D. thanks the support by 100 Talents Program of the Chinese Academy of Sciences. W.-L.W. thanks the support by Open Research Fund of State Key Laboratory of Precision Spectroscopy (East China Normal University).

\end{document}